\documentclass[]{trbunofficial}
\usepackage{graphicx}
\usepackage{pifont}
\usepackage[usenames, dvipsnames]{color}
\newcommand{\xmark}{\ding{55}}%

\usepackage[hidelinks]{hyperref}

\AuthorHeaders{A.~Haig, S.~A.~Hayati, and A.~Tomasic}
\title{Real-Time Detection of Crowded Buses via Mobile Phones}

\author{%
  \textbf{Alex Haig}\\
  Carnegie Mellon University\\
  adhaig@andrew.cmu.edu\\
  \hfill\break
  \textbf{Shirley Anugrah Hayati}\\
  Carnegie Mellon University\\
  shayati@cs.cmu.edu\\
  \hfill\break%
  \textbf{Anthony Tomasic}\\
  Carnegie Mellon University\\
  anthony.tomasic@gmail.com
}

\TotalWords{4727} 

\begin{document}
\maketitle

\section{Abstract}

Automated passenger counting (APC) technology is central to many aspects of the public transit experience. APC information informs public transit planners about utilization in a public transit system and operations about dynamic fluctuations in demand. Perhaps most importantly, APC information provides one metric to the rider experience -- standing during a long ride because of a crowded vehicle is an unpleasant experience. Several technologies have been successfully used for APC including light beam sensing and video image analysis. However, these technologies are expensive and must be installed in buses. In this paper, we analyze a new source of data using statistical models: rider smartphone accelerometers. Smartphones are ubiquitous in society and accelerometers have been shown to accurately model user states such as walking and sitting. We extend these models to use accelerometers to detect if the rider is standing or sitting on a bus. Standing riders are a signal that the bus is crowded. This paper provides evidence that user smartphones are a valid source of participatory sensing and thus a new source of automated passenger counting data.

\hfill\break%
\noindent\textit{Keywords}: Automatic Passenger Counter, Participatory Sensing, Machine Learning
\newpage

\section{Introduction}
Accurate knowledge about the utilization of public transit vehicles by riders, such as bus fullness, is critical information for public transit planners. Automatic Passenger Counters (APC) are used to count riders in public transit systems. APC information is used by transit planners to detect transit bottlenecks. A vehicle with a passenger counter higher than the numbers of seats available indicates that capacity is approaching the limit. APC counting is also used in real time by operations teams to assess overcrowding of vehicles -- a signal to dispatch additional vehicles on demand.  Finally, APC data provides evidence that the rider experience is poor - high APC counts, relative to the capacity of the vehicle, indicate that riders are probably standing in a crowded vehicle during the trip.

In this paper, we investigate a new technique for recording information about bus vehicle fullness. Participatory sensing, via a user's smartphone accerolmeter and GPS, can be used in transit systems to gather information about users and vehicles \cite{tomasic2015performance} without requiring action from the user -- in this case whether the rider is sitting or standing. This information, aggregated across riders in a vehicle, provides a signal about current conditions. The fullness of a bus can be estimated by the number of riders standing or sitting on the bus. Since smartphones are now equipped with accelerometers, machine learning algorithms can predict the state of riders and thus infer the state of the vehicle with respect to fullness. This method of determining fullness is thus complementary to APC data.

While APC data may be a good measure of rider counts in aggregate, it can suffer from high error when considering individual vehicles (cf.~Results Section) due to compounding errors. Thus, APC data is not always reliable. On the other hand, using phone sensors may not be able to produce an exact rider count, but aggregating across phone sensors provides an independent, direct, measurement of the state of riders. In addition, the error associated with each measurement is independent.

Third party transit information providers want information about the fullness of a bus prior to its arrival to provide to riders. Such information will be especially useful for riders with accessibility issues. However, these transit information providers typically do not have access to the APC data in real-time. Participatory sensing provides a direct method of providing this information, independent of the budgetary constraints of the transit agency \cite{steinfeld2011}.


\section{Proposed Architecture}

\begin{figure}[t]
    \caption{System Functional Architecture}
    \label{fig:arch}
    \centering
    \fbox{\includegraphics[width=0.9\textwidth,trim={2cm 2cm 2cm 5.5cm},clip]{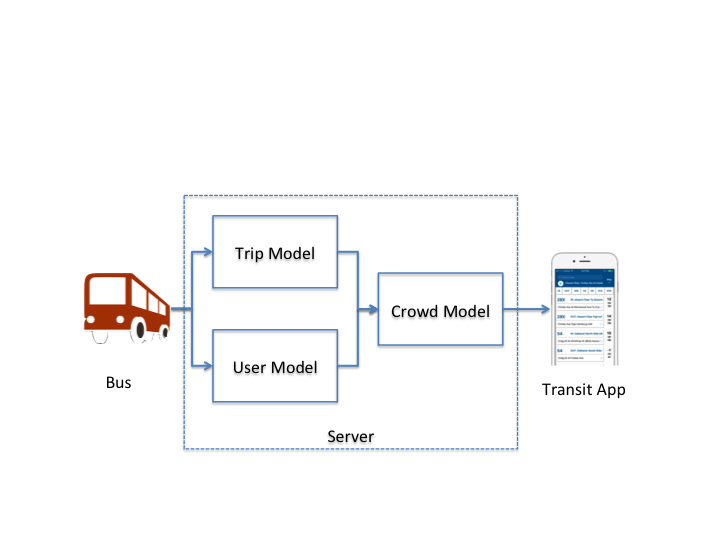}}
\end{figure}

The proposed architecture for the system consists of five components (Figure~\ref{fig:arch}). The bus symbol represents riders utilizing a mobile smartphone that gathers accelerometer and telemetry data. This information is transmitted to a server in the cloud that contains three models. The Trip Model clusters users into the same bus trip using prior research methods (\cite{zimmerman2011field}, \cite{Stenneth2011}). The User Model, detailed below, models the state of the rider. The Crowd Model, detailed below, translates the aggregation of the results of the user model for each trip to an indication of the state of the bus (e.g., many seats, few seats, crowded, very crowded). This information is transmitted to users utilizing a transit information application. For simplicity, in this architecture, all models are run on a cloud server. However, the user model can also be on the client phones of users on the bus. This client architecture reduces network communication costs.

\section{Methodology}
This study is organized into three linked sections. In the first section, we analyze APC data for the Port Authority of Allegheny County (PAAC) to estimate its accuracy. The next section describes the methodology, via direct field observations, for measuring the relationship between bus fullness and riders standing on the bus. In the third section, accelerometer data from smartphones is logged. The logs provide training data to model a rider's current state, specifically to detect if they were on a bus, and if so were they standing or sitting.

\subsection{Automatic Passenger Counter Data Accuracy}

To examine the APC data, we obtained two months worth of data for the PAAC bus system, from November of 2016 and March of 2017. PAAC buses use light beams mounted at the doors of buses to count riders entering and exiting buses. This data does not contain ground-truth measurements of rider counts. We estimate its accuracy by using an internal consistency argument: the total number of riders getting on a bus must equal the total number of riders getting off a bus for each bus's block. We use blocks because a rider can theoretically ride the entire chain of trips within a block continuously. However, a rider cannot get on the bus before the start of the block and must get off before or at the end of the block. The block error is thus the difference between the total number of counted embarking minus the total number of disembarking. This value should always be zero. Estimating the error in this case is just a simple matter of calculating the difference between the sum of the number of people recorded as getting on, and the sum of the number people recorded as getting off. Of course, this methodology does not account for errors that cancel each other out, such as person getting both on and off the bus without being recorded in either instance, but it still gives us a rough upper-bound on the accuracy overall.

\subsection{Bus Fullness Observation}
If someone is standing on the bus then there is a high probability that the bus is full or close to full. To test this hypothesis, we conducted field observations while riding buses. Every time a bus left a stop, we counted the number of riders sitting and the number of riders standing. In addition, we kept track of the bus number for each bus which uniquely identifies the bus model and thus the number of seats. 

Overall we recorded 246 data points from March to April 2018 across 12 different buses on 5 routes over 168 stops.

This gathered data serves as our Crowd Model. This model is a simple function that translates the empirically gathered data into predefined categories. 

\subsection{User Activity State Modeling}
Activity detection consists of multiple steps: gathering a supervised set of training data, iterative engineering of a machine learning model, and generation of the final evaluation statistics of the best model.

\subsubsection{Data Collection}
To determine the activity state of riders, we built a custom mobile app for Android and iOS that collected raw data from the GPS, accelerometer, and compass sensors in the phone. The app allows users to select one of the 15 activity conditions show in Table \ref{tab:statistics}. Upon selection of a state, the app waits 5 seconds to allow the user to position the phone correctly, records data for 15 seconds, and then stops automatically. This cycle comprises one data session. The timing for the recording of data for each session results in clean data that is less affected by random real-world factors. Using this app and 5 different phones we recorded a total of 3209 sessions (approximately 13 hours) of real-world data from October 2017 to July 2018. Table \ref{tab:statistics} shows the specific breakdown of the data including how the data was split into training and test sets when training our machine learning models. The true distribution of users' activity states varies per user in the real-world, so we balanced our test set across the different states in order to get a fair comparison of the accuracy for each state. The entire dataset was split into two parts: a training set and a test set. During training, a validation set was split from the training set. This set was used to iteratively improve the machine learning models described below. The test set was used after the end of the iterative improvements to produce the final values reported in this paper.

\begin{table}[h]
\caption{Dataset Statistics}
	\label{tab:statistics}
	\begin{center}
	    
\begin{tabular}{lll|rr}
\multicolumn{3}{c}{\textbf{User State}} & \multicolumn{2}{c}{\textbf{Sessions}} \\
\textbf{State}      & \textbf{Standing/Sitting} & \textbf{Phone Position}    & \textbf{Train} & \textbf{Test} \\ \hline \hline
Bus        & Standing         & Pocket   &  198     & 20     \\
           & Standing         & Hand     &  217     & 20     \\
           & Standing         & Backpack &  139     & 20     \\ \hline
Bus        & Sitting          & Pocket   &  212     & 20     \\
           & Sitting          & Hand     &  247     & 20     \\
           & Sitting          & Backpack &  210     & 20     \\ \hline
Walking    & \xmark           & Pocket   &  197     & 20     \\
           & \xmark           & Hand     &  194     & 20     \\
           &  \xmark          & Backpack &  196     & 20     \\ \hline
Stationary & Standing         & Pocket   &  202     & 20     \\
           & Standing         & Hand     &  187     & 20     \\
           & Standing         & Backpack &  165     & 20     \\ \hline
Stationary & Sitting          & Pocket   &  179     & 20     \\
           & Sitting          & Hand     &  183     & 20     \\
           & Sitting          & Backpack &  183     & 20    \\ \hline \hline
Total      &                  &          &  2909    & 300   \\ \hline  
    \end{tabular}
\end{center}
\end{table}

\subsubsection{Data Preprocessing}
To clean the noise in the dataset, we filter the dataset by selecting sessions that have at least 200 points of sensor readings over a time span of at least 10 seconds. This cleaning was performed to account for recording errors caused by bugs in the app itself. Such errors would be significantly less common in a fully implemented production system. Table \ref{tab:statistics} shows our data after cleaning.

\subsubsection{Features}
From the phone sensors, the app obtains seven values: acceleration in x-axis, acceleration in y-axis, acceleration in z-axis, latitude, longitude, speed and heading. Then, we follow \trbcite{elhoushi2016} for extracting statistical features from each of these raw sensor values -- except for heading which reduced the performance of our machine learning models. Note that we do not implement all the features mentioned in \trbcite{elhoushi2016}. The six values are translated into a set of features that produce high accuracy (cf.~Results Section) on the user states we are interested in. The total number of features used is 48: 8 summary statistics -- as described 
below -- for each of the 6 raw sensor values


\begin{itemize}
    \setlength\itemsep{.66em}
    \item Mean: \(\displaystyle \text{mean}(x) = \bar{x} = \frac{1}{N}\sum_{n=1}^{N} x_n\)
    \item Mean absolute value: \(\displaystyle \text{mean}(|x|) = |\bar{x}| \)
    \item Median: Middle value separating the higher half of data from the lower half of data
    \item Variance: \(\displaystyle \text{var}(x) = \sigma^2(x) = \frac{1}{N}\sum_{n=1}^{N} (x_n - \bar{x})^2 \)
    \item Standard Deviation: \(\displaystyle \text{std}(x) = \sigma(x) = \sqrt{\text{var}(x)} \)
    \item Average Absolute Difference: \(\displaystyle |\overline{x - \bar{x}}| \)
    \item Interquartile Range: Difference between 75th percentile and 25th percentile
    \item 75th Percentile: Value separating the higher 25\% of data from the lower 75\% of data
\end{itemize}

\subsubsection{Model Construction}
Using scikit-learn \cite{scikit-learn}, we trained several classifiers to predict the activity state of a user from among the 15 classes defined in Table \ref{tab:statistics}. The default parameters for the models were used unless otherwise specified. Three different classifiers were investigated: Random Forests, Support Vector Machines (SVM), and Multilayer Perceptrons (MLP). 


\textbf{Random forest} is an ensemble classifier that randomly constructs many decision trees which vote for the most popular class \cite{breiman2001}. Individual trees were limited to a maximum depth of 20 nodes. In decision tree learning, each leaf represent a class; each interior node represent a criteria on a feature that selects one branch from a node to its children (either another node or a leaf). A decision tree is iteratively built top down using the Gini impurity criterion to split the feature space. 


\textbf{Support vector machines} map non-linear input vectors into a higher dimensional feature space (\trbcite{Weston98}). For binary classification, a decision line is constructed to separate data points according to their labels and then the margin is maximized. 

\textbf{Multilayer perceptron} is a feed-forward neural network that consists of one or more hidden layers with nonlinear activation function \cite{Freund1999}. Our network has 1 hidden layer with 15 hidden units. 


\section{Results}
\label{sec:results}

\subsection{APC Data Accuracy}
\begin{figure}
    \caption{APC Block Errors for November, 2016 Dataset}
    \label{fig:apc-accuracy}
    \centering
    \includegraphics[width=\textwidth]{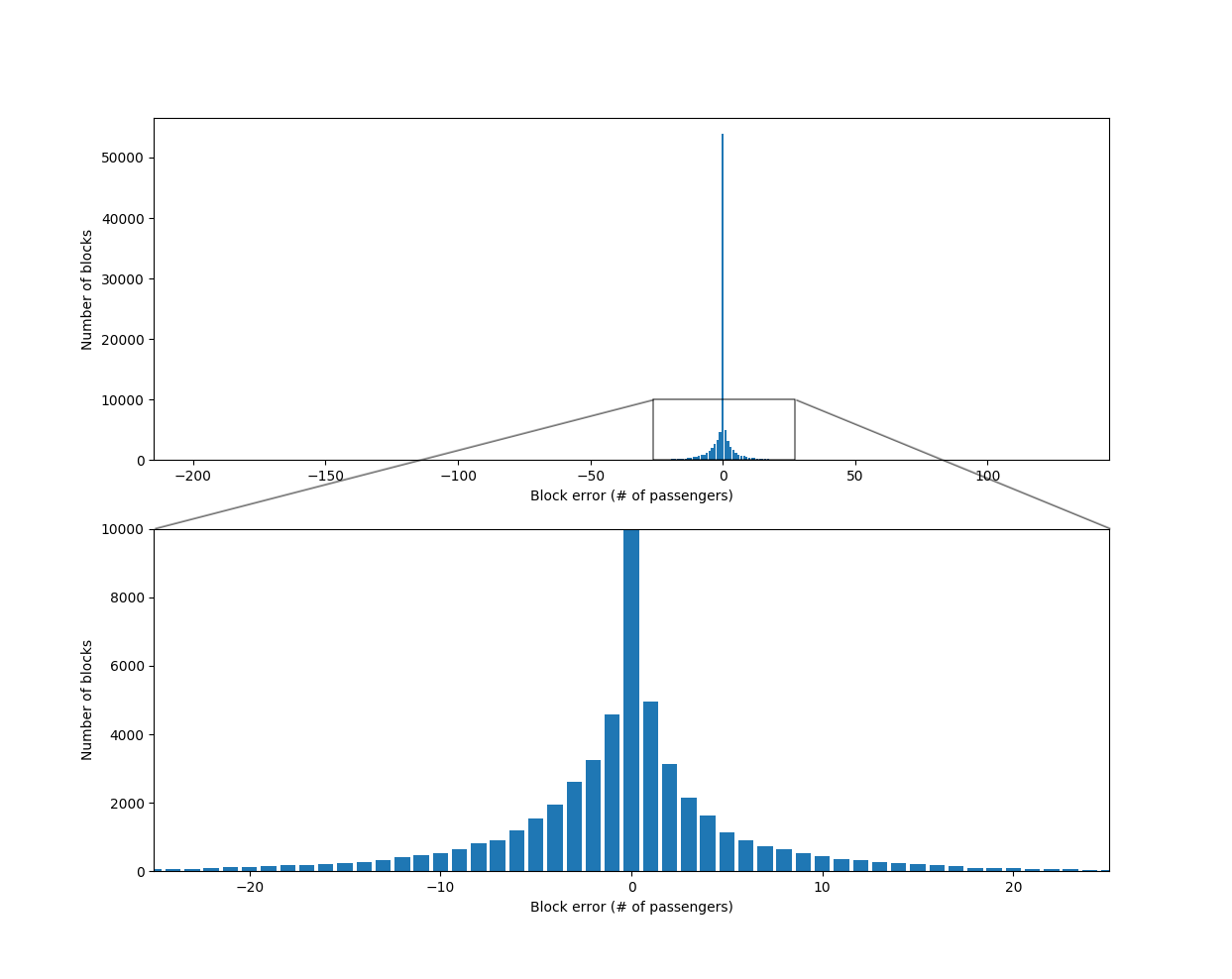}
\end{figure}

The APC data accuracy is measured by using internal consistency so the error represents over- or under-counting for a block. A histogram (Figure \ref{fig:apc-accuracy}) of the data indicates that the vast majority of the data values are zero -- see the top graph. The bottom graph, cropped and magnified, indicates that measurement error is balanced about the mean of zero. Standard statistics of the error (Table  \ref{tab:apc-accuracy}) confirms this analysis. 
 
Overall we found that the mean error is near zero for both the 11/16 and 03/17 datasets, but the variance is quite high. The minimum error was -198 and the maximum was 188 for individual blocks.  These statistics indicates that APC data is accurate for measuring the number of riders in aggregate, but with 43.6\% of blocks containing errors, the reliability of APC counts needs improvement before it is used to measure the number of riders on buses in real-time.

\begin{table}
	\caption{APC Block Errors Sample Statistics}
	\label{tab:apc-accuracy}
	\begin{center}
		\begin{tabular}{ll r}
			\textbf{Dataset} & \textbf{Measurement} & \textbf{Result} \\\hline \hline
			11/16 &  Size (Blocks)  &  94,498 \\
			      &  Mean           & -0.24 \\
			      &  Variance       & 34.61 \\
			      &  Standard Error of Mean & 0.019 \\ 
			      &  Nonzero        & 43\% \\ \hline
			03/17 &  Size (Blocks)  & 77,797 \\
			      &  Mean           & -0.09 \\
			      &  Variance       & 45.56 \\
			      &  Standard Error of Mean & 0.024 \\
			      &  Nonzero        & 44\% \\	\hline \hline
		\end{tabular}
	\end{center}
\end{table}

\subsection{Bus Fullness, Standing and Sitting}
\begin{figure}[h]
    \caption{Riders Standing vs.~Seats Taken}
    \label{fig:stand_v_full}
    \centering
    \includegraphics[width=\textwidth]{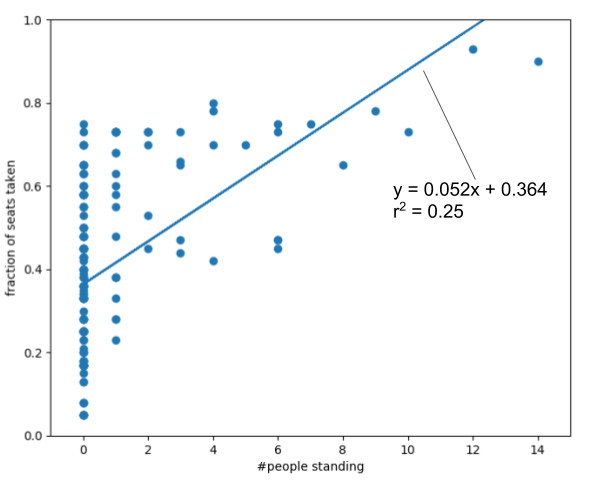}
\end{figure}

To understand the behavior of riders on buses, we performed field observations on several routes and buses. Generally people preferred sitting on buses when seating was easily available. Anecdotal evidence provides some few exceptions to this rule. Riders stood when seating was difficult to reach, the rider was on the bus for a short period, or in special circumstances, such as a parent with a child in a stroller. Further research would investigate the behaviour of these riders through interviews to better account for them in our model.
A regression analysis of the data (Figure \ref{fig:stand_v_full}) shows a clear relationship between the number of riders standing on the bus and the fullness of the bus. There is a clear upwards trend showing that the number of riders standing is indeed predictive of the fullness of the bus. Surprisingly, this behavior becomes apparent when the seats available on a bus reaches 50\% or so. We had informally expected that the behavior would occur only when most of the seats were taken. An interesting future analysis would interview riders to discuss the conditions under which riders perceive a bus as crowded.




\subsection{Activity Detection}
The results in this section are reported on the approximately 10\% of the recorded data reserved for testing.
Gathering 15 user states insured that the activity detection training set contains a robust set of different circumstances of mobile phone user for riders. Our iterative investigation of various machine learning models produced somewhat low performance (Table \ref{tab:all}).
The highest accuracy (79.33\%) is achieved by using Random Forest.

\begin{figure}[h]
    \caption{Random Forest Confusion Matrix for All Classes}
    \label{fig:conf-mat-all}
    \centering
    \includegraphics[width=\textwidth]{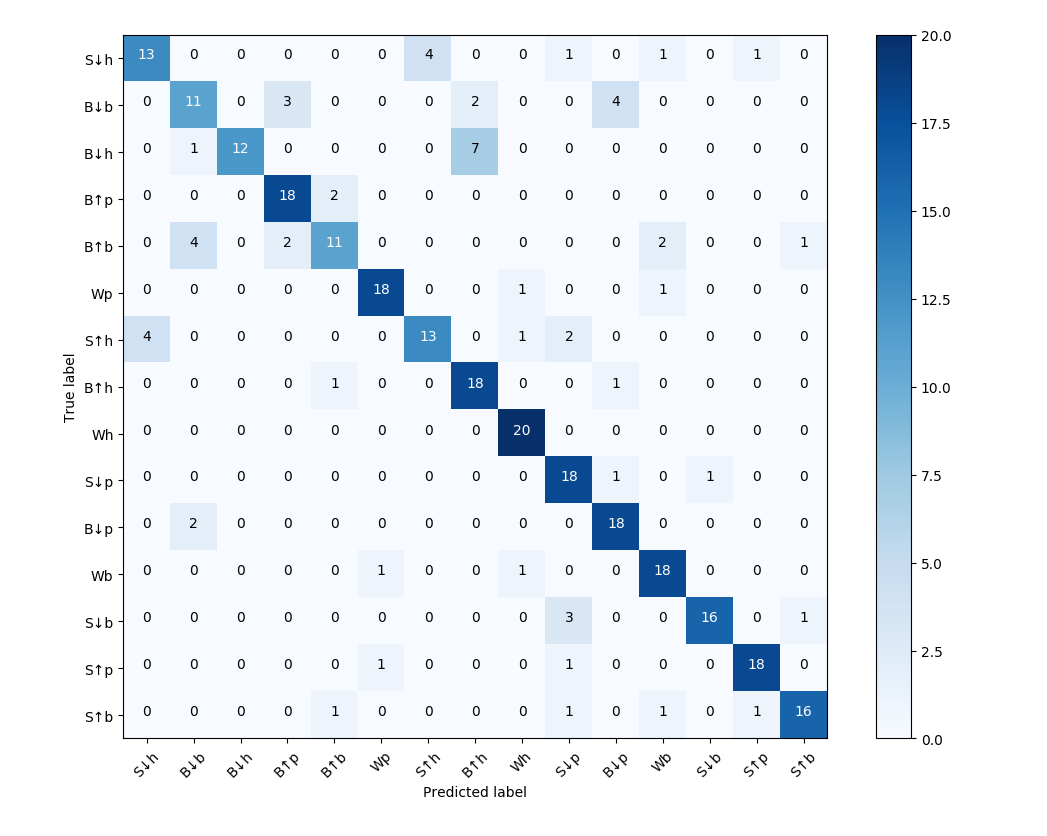}
\end{figure}

The confusion matrix of the Random Forest Model (Figure \ref{fig:conf-mat-all}) categorizes the errors that the model makes as counts off the diagonal. Table~\ref{tab:conf-mat-key} is the key to the axes of the confusion matrix. All 20 test examples with the state of ``walking hand'' are predicted correctly. Another interesting finding is that 7 examples with the state ``on the bus, sitting, hand'' are predicted as ``on the bus, standing, hand'' although 18 examples with the state ``on the bus, standing, hand'' are predicted correctly. Another mistake that our model often makes is distinguishing the location of the phone (hand, pocket, or backpack). For instance, 4 examples of ``on the bus, sitting, backpack'' state are predicted as ``on the bus, sitting, pocket.'' While the Random Forest model can distinguish the ``on the bus'' state from the ``off the bus'' state, it sometimes makes mistakes discriminating ``standing'' and ``sitting''. For instance, 4 examples of ``on the bus, standing, backpack'' are predicted as ``on the bus, sitting, backpack''.

\begin{table}[h]
	\caption{Performance of Model using All Classes}
	\label{tab:all}
	\begin{center}
		\begin{tabular}{l r}
			\textbf{Algorithm} & \textbf{Accuracy} \% \\\hline \hline
			Random Forest   & 79.33 \\
			SVM             & 56.17 \\
			MLP             & 72.65 \\\hline \hline
		\end{tabular}
	\end{center}
\end{table}

\begin{table}[h]
	\caption{Key for confusion matrix classes}
	\label{tab:conf-mat-key}
	\begin{center}
		\begin{tabular}{l r}
			\textbf{Symbol} & \textbf{Meaning} \\\hline \hline
			S & Stationary \\
			W & Walking \\
			B & On Bus \\
			h & Hand \\
			p & Pocket \\
			b & Backpack/Bag \\
			$\uparrow$ & Standing \\
			$\downarrow$ & Sitting \\
			\hline \hline
		\end{tabular}
	\end{center}
\end{table}

\begin{figure}[h]
    \caption{Confusion matrix for Stationary, Walking, On Bus}
    \label{fig:conf-mat-transport}
    \centering
    \includegraphics[width=0.5\textwidth]{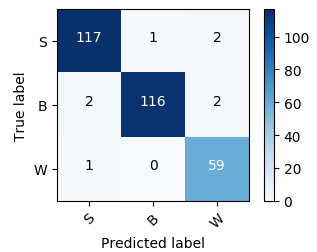}
\end{figure}

Since the goal of our work is to detect if the rider is on the bus, and if so whether they are standing or sitting, we conducted two experiments with merged classes.  The first experiment distinguishes the user's transportation mode by merging the 15 classes into three classes: ``on bus'', ``walking'', and ``stationary''. The confusion matrix for this experiment is shown in Figure \ref{fig:conf-mat-transport}. An analysis of the matrix shows very good performance with proportionately few errors.

\begin{figure}[h]
    \caption{Confusion matrix for On Bus Sitting, On Bus Standing, Other}
    \label{fig:bus-sit-stand-other}
    \centering
    \includegraphics[scale=1]{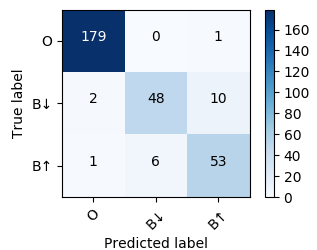}
\end{figure}

A second experiment discriminates the user's state on the bus and off the bus, and if the user is sitting or standing on the bus by merging classes into: ``standing on the bus'', ``sitting on the bus'', and ``other''. Again we get good performance as shown in Figure \ref{fig:bus-sit-stand-other}. Table \ref{tab:merged} summarizes the performance of our models on both of these merged classes experiments. Both experiments have high accuracy.

These experiments show that even though the performance of the activity model on the full set of states is fairly low compared to previous work \cite{elhoushi2016}, the model still performs very well on detecting the user states we are most concerned with. The model very accurately determines whether riders are standing on a bus. Given that the model is operating on many riders on a bus (such as part of a smartphone operating system), aggregating the results of these models produces an estimate of the number of people standing on a bus. Given this value, the number of seats taken can be estimated by cross referencing the data in Figure~\ref{fig:stand_v_full}.

\begin{table}[h]
	\caption{Performance of Model using Merged Classes}
	\label{tab:merged}
	\begin{center}
		\begin{tabular}{l r}
			\textbf{Classes} & \textbf{Accuracy} \% \\\hline \hline
			Bus, Walking, Stationary              & 97.33 \\
			Bus Standing, On Bus Sitting, Other   & 93.33 \\\hline \hline
		\end{tabular}
	\end{center}
\end{table}

\section{Discussion}
The following discussion examines how our results relate to prior work. It is organized into sections on automated passenger counting, user activity sensing by mobile devices, mobile transit information systems, and participatory sensing.

\subsection{Automated Passenger Counting}

Although we did not use ground-truth data, our results on the errors found within APC data are consistent with those found in previous work that did use ground-truth data, \trbcite{kimpel2003automatic}. Using cameras to measure actual bus loads, they found that the difference in load between APC counts and ground-truth was 6.08\%. Additionally, they used reverse regression to measure APC precision and found a bias of 8.81\%. Both results were found to be statistically significant. 

One other significant difference between the analysis in \trbcite{kimpel2003automatic} and our own is that they performed significant cleaning on the data before measuring the APC accuracy by removing approximately 35\% of their data. We did not perform such cleaning when examining the APC data because postprocessing the data in such a way would not be possible in a real-time system.

Another method of counting riders (\trbcite{chen2008people}) is to install cameras in buses and use computer vision techniques to detect riders getting on or off. Such a system can achieve an accuracy comparable to that found in APC systems, but it still requires bus agencies to install cameras in all of the buses, potentially a significant expense. 

\subsection{User Activity Sensing in Mobile Phones}
Many prior works employ mobile phone sensors and supervised learning to determine user activity, either related to transportation mode or not. In detecting user activity, previous studies explored three main approaches: location-based (GPS), sensor-based (accelerometer), and hybrid of location-based and sensor-based approaches. 

\trbcite{Ashbrook2003}, \trbcite{zheng2008}, \trbcite{Zheng2010}, and \trbcite{Stenneth2011} use GPS to infer the motion or transportation mode of users. \trbcite{Sohn2006} explored the usage of GSM sensors to detect daily physical activities such as walking, driving, or being stationary. \trbcite{myrvoll2017counting} counts the number of passengers on a bus through monitoring WiFi probe requests, a potential complementary source of passenger count data.
 
Microelectromechanical systems motion sensors (accelerometer, gyroscope, magnetometer, and barometer), without any location signals, can be used to detect indoor motion, such as walking, stationary, and going up/down stairs (\trbcite{elhoushi2016}). 
\trbcite{nham2008} collected acceleration data from mobile phones for predicting walking, biking, running, driving, riding a train, or riding a bus. \trbcite{Hemminki2013} introduce a gravity estimation technique for accelerometers to produce more accurate horizontal acceleration measurements to discern information from user motion from noise like gravity. 
The user model in this paper extends this body of work to the specific case of classifying user activities on a bus.

Several works focus on the transportation mode of a user.
\trbcite{fang2016} classified transportation mode using the accelerometer, magnetometer, and gyroscope sensors from smartphones and ran three machine learning algorithms (decision trees, K-nearest neighbor, and support vector machine). \trbcite{Yu2014} applied four strategies which are big data, small footprint, data substitution, and multi-tier design in minimizing power consumption and memory requirement to detect transportation mode. \trbcite{Reddy2010} used both accelerometer and GPS to identify transportation mode (stationary, walking, running, biking, in motorized transport). 
\trbcite{Jahangiri2015} tried to detect the transportation mode (including driving a car, riding a bicycle, riding a bus, walking, and running) of a user via mobile phones using K-nearest neighbor, support vector machines (SVMs), and tree-based models such as decision tree, bagging, and random forest. Their best result was achieved by using random forest and SVM, similar to our results. 

\subsection{Mobile Transit Information Systems}

This paper also extends a body of previous work using mobile phones to collect transit data from users and then using this data to improve the users experience on transit systems. For example the Tiramisu Transit system \cite{zimmerman2011field, tomasic2015performance} used crowd sourcing to generate real-time arrival information for the PAAC bus system before real-time information was provided by PAAC itself. The system presented in this study would be a novel extension by using user's phones to detect real time bus fullness, and then improving the user's transit experience by presenting fullness information.  \trbcite{zimmerman2011field} provides methodology for clustering users into the associated bus trips that they are on. The system simply asks the user directly as a form of crowdsourcing. Incentives to encourage users to report this information are studied in \trbcite{tomasic2014motivating}. \trbcite{Stenneth2011} uses automated methods to gather the same information.

\subsection{Participatory Sensing}
Our work is also related to participatory sensing for applications related to public transportation tracking.  \trbcite{zhou2012} presented a crowdsourced bus arrival time estimation system by utilizing low-energy sensing resources, such as cell tower signals, movement statuses, and audio recordings. \trbcite{Thiagarajan2010} also proposed crowdsourced transit (bus and train) tracking. They introduced power-efficient activity classification to detect if riders are in the vehicle or not, low-memory route-matching algorithm, and a method to track underground trains. 

\trbcite{pi2018} studied the how users perceived bus fullness by investigating the relationship between APC data and crowdsourced bus fullness ratings from users of Tiramisu. They built a classification model to infer riders' fullness perception from APC data and other factors. Their model could also be used by transit agency to help predict bus fullness. Our work is distinguished by the fact that we indirectly but independently estimate bus fullness using phone sensors whereas they use APC data.



\section{Conclusion}

Real time bus fullness (the crowd on a bus) information is important to both transit users and planners. In this paper we showed that the reliability of APC data needs to be improved before presenting this information in real time.
We then conjectured that the same information could be derived by determining how many people were standing in a bus. Field observations confirmed a direct relationship between people standing and the number of seats taken on a bus. To make this information available to riders in real time, the paper demonstrated that smartphone sensors can provide relevant real time data. Several machine learning models were applied to the logs of users' activities, both on and off buses, and were shown to be highly accurate in detecting when a user was standing or sitting on a bus. Thus, adding these models to mobile phone operating systems or applications, and transmitting the results of these models in real-time, would provide an accurate foundation for this feature in transit information systems.

In future work, the results of this work can easily be extended to many different public transit systems, such as trains or subways. Additionally, with more data we could improve the results of our activity model and better learn the relationship between bus fullness and the number of people standing. 






\section{Acknowledgements}

Special thanks to the software engineers of the first version of the data collection app: Napat Luevisadpaibul and Wenqing Yuan. This work was supported in part by the National Institute on Disability, Independent Living, and Rehabilitation Research (NIDILRR Grant 90RE5011-01-00) and by US DOT FAST Act - Mobility National (2016 - 2022) - CMU 2017 Mobility21 UTC \#31.

\section{Author Contribution Statement}

The authors confirm contribution to the paper as follows: study conception and design: Alex Haig, Shirley Anugrah Hayati, Anthony Tomasic; data collection: Alex Haig, Shirley Anugrah Hayati, Anthony Tomasic; analysis and interpretation of results: Alex Haig, Shirley Anugrah Hayati, Anthony Tomasic; draft manuscript preparation: Alex Haig, Shirley Anugrah Hayati, Anthony Tomasic. All authors reviewed the results and approved the final version of the manuscript.

\newpage

\bibliographystyle{trb}
\bibliography{main}
\end{document}